\documentclass[11pt]{article}

\usepackage{geometry}
\usepackage{amsmath,amssymb}
\usepackage{natbib}
\usepackage{booktabs}
\usepackage{graphicx}
\usepackage[colorlinks,citecolor=green,linkcolor=blue,bookmarks=false,hypertexnames=true]{hyperref}

\renewcommand{\cite}{\citep}
\newcommand{\mailto}[1]{\href{mailto:#1}{\texttt{#1}}}
\renewcommand{\toprule}{%
  \noalign{\vskip\abovetopsep}%
  \hline\hline
  \noalign{\vskip\belowrulesep}%
}
\renewcommand{\midrule}{%
  \noalign{\vskip\aboverulesep}%
  \hline
  \noalign{\vskip\belowrulesep}%
}
\renewcommand{\bottomrule}{%
  \noalign{\vskip\aboverulesep}%
  \hline
  \noalign{\vskip\belowbottomsep}%
}

\title{Predictability of El Ni\~no from Delayed Observations}
\author{
Francisco J. Beron-Vera\\
Department of Atmospheric Sciences\\
Rosenstiel School of Marine, Atmospheric \& Earth Science\\
University of Miami\\
Miami, FL 33149, USA\\
\mailto{fberon@miami.edu}
}
\date{Started: August 6, 2026.  This version: \today.}

\begin{document}

\maketitle

\begin{abstract}
    Using monthly Ni\~no-3.4 anomalies through July 2026, we investigate how much predictive information is contained in delayed observations of the index. Ridge regression identifies informative delays, while multilayer perceptron and sparse identification of nonlinear dynamics (SINDy) models test whether nonlinear complexity provides additional direct forecast skill; gated recurrent unit (GRU) and long short-term memory (LSTM) networks provide a complementary test in which the temporal representation is learned internally. Delayed observations substantially improve forecasts over persistence and climatology at leads of up to six months, but increasing model complexity provides no systematic improvement. Historical recursive experiments favor a simple explicit SINDy recurrence and select shallow recurrent architectures, with no appreciable gain from learning the temporal representation internally. These results support a compact predictive representation of Ni\~no-3.4 evolution in which the representation of past information is more consequential than model complexity. As a prospective application, the selected models are used to forecast the developing 2026 event beyond the last available observation and to compare its predicted evolution with completed historical El Ni\~no events.
\end{abstract}

\section*{Plain Language Summary}

El Ni\~no prediction often draws on many oceanic and atmospheric variables. This study examines how much predictive information is already contained in the past history of the Ni\~no-3.4 index. A small number of delayed observations substantially improves forecasts, while more complex nonlinear and recurrent models provide little additional skill. The results indicate that representing the recent history of Ni\~no-3.4 is more important for prediction than increasing model complexity.

\section*{Key Points}

\begin{itemize}
    \item A small number of delayed Ni\~no-3.4 observations provides substantial forecast skill relative to persistence and climatology.
    \item Nonlinear MLP and SINDy models provide little improvement over linear ridge regression once informative delayed coordinates are included.
    \item GRU and LSTM networks that learn temporal representations internally provide no gain over a simple delayed-coordinate recurrence.
\end{itemize}

\section{Introduction}

The El Ni\~no--Southern Oscillation (ENSO) is the dominant mode of interannual tropical climate variability and a major source of predictability for climate anomalies worldwide \cite{Dijkstra-Burgers-02, Clarke-14}. Its evolution reflects coupled ocean--atmosphere feedbacks operating over a range of time scales. In particular, conceptual models of ENSO have long emphasized memory and delayed negative feedback. The delayed-oscillator picture represents the influence of basin-scale equatorial-wave adjustment through an explicit time delay \cite{Suarez-Schopf-88, Battisti-Hirst-89}, whereas recharge--discharge descriptions emphasize the slower evolution of equatorial upper-ocean heat content \cite{Jin-97}. Despite their different physical interpretations, both imply that the present sea-surface-temperature anomaly alone need not contain all the information relevant to its subsequent evolution.

The role of memory in ENSO motivates an examination of the predictive information contained in past values of the observed Ni\~no-3.4 sea-surface-temperature anomaly. Delayed observations of a scalar variable can, under appropriate conditions, provide coordinates for an underlying dynamical state \cite{Takens-81}. We do not assume that a finite set of Ni\~no-3.4 delays reconstructs the ENSO state, nor do we identify individual delays with particular oceanic adjustment processes. Our aim is to determine empirically whether a small collection of past Ni\~no-3.4 anomalies provides useful predictive coordinates, what memory ranges emerge, and whether forecast skill saturates as additional delayed observations are included.

Data-driven methods have become increasingly prominent in ENSO prediction \cite{Dijkstra-etal-19,Wang-etal-23}. Much of this work seeks long-lead forecast skill by exploiting spatially distributed oceanic and atmospheric information. Convolutional neural networks trained on tropical Pacific sea-surface-temperature fields, supplemented by climate-model simulations, have demonstrated skill at long leads \cite{Ham-etal-19}, while more recent work has used deep learning to identify the oceanic information underlying such predictability \cite{Chen-etal-25}. We use only the monthly Ni\~no-3.4 index to examine how much of its predictability can be recovered from its own delayed history.

Following the delayed-coordinate forecasting framework introduced by Beron-Vera et al.~\cite{Beron-etal-26-NODY}, ridge regression \cite{Hoerl-Kennard-70} is used to identify forecast-relevant delayed coordinates, while multilayer perceptron (MLP) \cite{Rumelhart-etal-86} and sparse identification of nonlinear dynamics (SINDy) \cite{Brunton-etal-16} models test whether nonlinear maps extract additional information from those coordinates. Gated recurrent unit (GRU) \cite{Cho-etal-14} and long short-term memory (LSTM) \cite{Hochreiter-Schmidhuber-97} networks broaden the comparison by learning the temporal representation internally from a consecutive sequence rather than using explicitly selected delayed coordinates.

This comparison separates two sources of forecasting complexity: the representation of the state and the map acting on that representation. It also allows an explicit evolution law to be sought with SINDy rather than treating forecasting solely as a black-box prediction problem. Using monthly Ni\~no-3.4 through July 2026, we find that forecast skill saturates with a small number of delayed coordinates and that nonlinear and recurrent complexity provides no systematic improvement. Historical El Ni\~no experiments favor a simple explicit SINDy recurrence, while pseudo-real-time hindcasts select shallow recurrent architectures. The selected models are then applied prospectively to the developing 2026 event.

\section{Methods}
\label{sec:methods}

We use the monthly Ni\~no-3.4 sea-surface-temperature anomaly from the NOAA Climate Prediction Center, based on ERSSTv6 \cite{Huang-etal-25a,Huang-etal-25b}, denoted by \(N(t)\), from December 1949 through July 2026. The index is the sea-surface-temperature anomaly averaged over \(5^\circ{\rm S}\)--\(5^\circ{\rm N}\), \(170^\circ{\rm W}\)--\(120^\circ{\rm W}\), with centered climatological base periods. We work directly with this monthly index, which provides the monthly input to the Oceanic Ni\~no Index (ONI), rather than ONI itself, whose overlapping three-month averages introduce temporal smoothing before delayed observations are considered. For a \emph{forecast lead} \(\Delta\), we seek an empirical predictive map
\begin{equation}
    \widehat N(t+\Delta)
    =
    F_\Delta\bigl(\mathbf z(t;\mathcal T);\theta\bigr),
    \qquad
    \mathbf z(t;\mathcal T)
    =
    \bigl(N(t-\tau_1),\ldots,N(t-\tau_m)\bigr),
    \label{eq:forecast}
\end{equation}
where \(\mathcal T=\{\tau_1,\ldots,\tau_m\}\), with \(\tau_1=0\), specifies the \emph{delayed coordinates} and \(\theta\) denotes the model parameters. The use of delayed coordinates is motivated by state-reconstruction ideas, but the existence of a predictive map for the particular finite lag sets considered here is not implied by Takens' embedding theorem \cite{Takens-81}. Whether these delayed coordinates provide useful predictive information is instead assessed empirically.

We first use ridge regression both as a forecasting model and to identify informative delayed coordinates. Candidate delays are drawn from \(\{0,1,2,3,4,6,9,12,15\}\) months, with the instantaneous coordinate always included. For each number \(m\) of coordinates, all admissible combinations of the remaining delays are considered, and the best-performing lag set is retained. We then use the ridge-selected coordinates to test whether nonlinear dependence provides additional predictive information with MLP and SINDy. The corresponding realizations of \eqref{eq:forecast} may be summarized as
\begin{equation}
\begin{split}
    F_\Delta^{\rm ridge}(\mathbf z)
        &= \alpha_0+\boldsymbol{\alpha}\cdot\mathbf z,\\
    F_\Delta^{\rm MLP}(\mathbf z)
        &= \mathbf w_2\cdot
        \sigma(\mathbf W_1\mathbf z+\mathbf b_1)+b_2,\\
    F_\Delta^{\rm SINDy}(\mathbf z)
        &= \boldsymbol{\Theta}(\mathbf z)\boldsymbol{\xi},
\end{split}
\label{eq:models}
\end{equation}
where the ridge coefficients are obtained by \(L_2\)-regularized least squares; \(\mathbf W_1,\mathbf w_2\) and \(\mathbf b_1,b_2\) are the MLP weights and biases, respectively, and \(\sigma\) is a nonlinear activation function; and \(\boldsymbol{\Theta}\) is a library of candidate polynomial functions whose coefficient vector \(\boldsymbol{\xi}\) is constrained to be sparse. The MLP expression in \eqref{eq:models} illustrates a single hidden layer, with deeper networks obtained by composing additional layers. Polynomial SINDy libraries through degree three are considered. Thus these three approaches act on the same explicitly prescribed delayed representation while ranging from linear to flexible and sparse nonlinear forecast maps.

We also consider GRU and LSTM networks, which instead learn an internal representation of temporal dependence from a consecutive sequence of past observations. Writing
\begin{equation}
    \mathbf x_t
        = \mathbf z(t;\{0,\delta,\ldots,(q-1)\delta\})
        = (N(t),N(t-\delta),\ldots,N(t-(q-1)\delta)),
    \label{eq:sequence}
\end{equation}
with \(\delta=1\) month, for an input sequence of \(q\) consecutive observations, their forecasts may be represented schematically as
\begin{equation}
    \mathbf h_t^R
        = \mathcal R_R(\mathbf x_t;\theta_R),
    \qquad
    \widehat N(t+\Delta)
        = F_\Delta^R(\mathbf x_t)
        = \mathbf w_R\cdot\mathbf h_t^R+b_R,
    \qquad
    R=\mathrm{GRU}\ \text{or}\ \mathrm{LSTM},
    \label{eq:recurrent}
\end{equation}
where \(\mathcal R_R\) denotes the recurrent processing of the input sequence, \(\mathbf h_t^R\) is the resulting hidden state, and \(\theta_R\) collects the learned weights and biases of the recurrent network. Both architectures carry information from one observation to the next through their internal states. A GRU uses update and reset gates, whereas an LSTM additionally maintains a cell state controlled by input, forget, and output gates. Thus, unlike ridge, MLP, and SINDy, for which temporal information is supplied explicitly through selected delays, GRU and LSTM learn a representation of the past internally through recurrence. We consider \emph{shallow} networks with one recurrent layer and, as a robustness test, \emph{deep} (i.e., stacked) architectures of increasing depth.

For ridge regression and SINDy, the record is divided chronologically into 70\% training and 30\% test samples. Neural-network models use a 70\%/15\%/15\% training/validation/test split, with the validation interval used for early stopping and model selection. Forecast skill is measured relative to persistence and zero-anomaly climatology according to
\begin{equation}
    R_p=\frac{\mathrm{RMSE}_{\rm model}}
              {\mathrm{RMSE}_{\rm persistence}},
    \qquad
    R_c=\frac{\mathrm{RMSE}_{\rm model}}
              {\mathrm{RMSE}_{\rm climatology}},
    \label{eq:skill}
\end{equation}
so that values below unity indicate improvement over the corresponding reference. Persistence provides the natural short-lead benchmark, although its skill deteriorates with lead. Zero-anomaly climatology instead provides a fixed reference whose RMSE measures the typical amplitude of the observed anomalies, so that \(R_c<1\) indicates model errors smaller than this intrinsic variability. Neural-network results are summarized across independent random initializations rather than selecting individual realizations according to test performance.

Direct forecasts in the ``open-loop'' configuration use observed delayed coordinates independently at each target time. To forecast beyond the final observation in July 2026, we instead iterate the one-month models recursively in ``closed loop,'' replacing unavailable observations in the delayed state by previous model predictions. Models with similar direct skill can behave differently under iteration, so recursive behavior is examined using completed historical El Ni\~no events, defined by \(N(t)\geq0.5\,^\circ\mathrm{C}\) for at least five consecutive months. For SINDy, these historical hindcasts provide a robustness diagnostic among polynomial degrees having comparable direct skill and motivate the degree-one map used for recursion.

For GRU and LSTM, architecture is selected through pseudo-real-time hindcasts. At each historical El Ni\~no event, the network is trained on all admissible input sequences available up to the forecast origin, with the final 15\% of these data reserved for validation and early stopping, and is then iterated recursively over the subsequent three months. The input sequence length is fixed, but the number of training sequences increases for later historical events as the available record expands. Each hindcast nevertheless remains out of sample relative to the data used to train the corresponding network. The three-month hindcast errors are combined across historical events using weights that decrease with the delayed-state distance between each forecast origin and the July 2026 initialization. The resulting similarity-weighted RMSE favors architectures that predict historical events accurately while giving greater weight to states resembling the current initialization; the architecture minimizing this RMSE is selected. After architecture selection, the chosen networks are retrained using the full record available through July 2026, with the same validation procedure, before producing the current recursive forecast.

\section{Results}
\label{sec:results}

Forecast skill improves rapidly as delayed observations are added and then saturates with a small number of coordinates. The best ridge models at forecast leads \(\Delta=1,3,\) and \(6\) months contain 4, 6, and 5 coordinates, respectively (Table~\ref{tab:direct}), with lag sets \(\{0,1,3,4\}\), \(\{0,3,6,9,12,15\}\), and \(\{0,1,3,9,15\}\) months. Short delays of a few months contain substantial forecast-relevant information, while delays of 9--15 months provide additional information at the longer leads. Beyond a few coordinates, however, the improvement is small. As forecast lead increases, improvement over persistence remains substantial (\(R_p<1\)), whereas skill relative to zero-anomaly climatology (\(R_c<1\)) progressively decreases. At 9--12 months the forecasts continue to outperform persistence but provide little improvement over climatology, with \(R_c\) close to unity. We therefore focus below on leads through six months, for which the delayed representation retains appreciable skill relative to both references.

\begin{table}[t]
\centering
\begin{tabular}{c c c cc cc}
\toprule
\(\Delta\) & \(m\) & \(\mathcal T\) &
\multicolumn{2}{c}{Ridge} &
\multicolumn{2}{c}{MLP} \\
{[months]} && {[months]} &
\(R_p\) & \(R_c\) &
\(R_p\) & \(R_c\) \\
\midrule
1 & 4 & \(\{0,1,3,4\}\)
  & 0.875 & 0.245 & 0.898 & 0.252\\
3 & 6 & \(\{0,3,6,9,12,15\}\)
  & 0.805 & 0.532 & 0.808 & 0.533\\
6 & 5 & \(\{0,1,3,9,15\}\)
  & 0.757 & 0.821 & 0.756 & 0.819\\
\bottomrule
\end{tabular}
\caption{Direct forecast skill using ridge-selected delayed coordinates. The number of coordinates \(m\) and lag set \(\mathcal T\) are selected by ridge regression. MLP results are medians over ten independent initializations.}
\label{tab:direct}
\end{table}

Replacing the linear ridge map by an MLP acting on the same coordinates does not systematically improve the forecast (Table~\ref{tab:direct}). In contrast, an instantaneous MLP using only \(N(t)\) gives \(R_p=1.009\), \(0.958\), and \(0.863\) at forecast leads \(\Delta=1,3,\) and \(6\) months, respectively. Thus the substantial gain comes from the delayed representation rather than from nonlinear flexibility of the forecast map.

SINDy allows polynomial interactions among the delayed coordinates while retaining a sparse explicit map. Direct SINDy forecasts outperform persistence and climatology at forecast leads \(\Delta=1,3,\) and \(6\) months, but provide no systematic evidence that greater nonlinear complexity improves forecast skill. For the \(\Delta=1\) month forecast, which is subsequently used for recursive prediction, the ridge-selected delayed state is
\begin{equation}
    \mathbf z(t)
    =
    \bigl(N(t),N(t-1),N(t-3),N(t-4)\bigr),
    \label{eq:selected-state}
\end{equation}
where, hereafter, integer time shifts are understood to be in months. The consistency of ridge, MLP, and SINDy indicates that much of the useful predictive information resides in a compact delayed representation of the scalar record rather than in the complexity of the map acting on it.

Recursive forecasting places a stronger constraint on model complexity. For SINDy, polynomial degrees one, two, and three have similar direct skill at forecast lead \(\Delta=1\) month,
\[
    R_p=0.892,\qquad 0.897,\qquad 0.916,
\]
respectively, so direct skill alone provides little basis for selecting among them. Their behavior under recursive iteration, however, is markedly different. Similarity-weighted three-month RMSEs over the completed historical El Ni\~no events are \(0.514\,^\circ{\rm C}\), \(125.6\,^\circ{\rm C}\), and \(12.3\,^\circ{\rm C}\) for degrees one, two, and three, respectively. The large errors of the nonlinear maps result from occasional recursive instabilities; for example, a degree-two hindcast initialized in October 1951 evolves from \(1.21\,^\circ{\rm C}\) to \(3.00\), \(17.82\), and \(1032.42\,^\circ{\rm C}\) over the subsequent three months. The historical recursive robustness test therefore favors \(p=1\).

Refitting the selected structure using all observations through July 2026 gives the explicit \(\Delta=1\) month map
\begin{equation}
\begin{split}
    \widehat N(t+1)={}&
    0.004667
    +1.082946\,N(t)
    -0.060292\,N(t-1)\\
    &-0.027001\,N(t-3)
    -0.103965\,N(t-4).
\end{split}
\label{eq:sindy-map}
\end{equation}
Thus, although polynomial libraries through degree three were allowed, the model favored for recursive forecasting is a linear recurrence involving only the four delayed coordinates identified above.

\begin{figure}[t]
    \centering
    \includegraphics[width=\columnwidth]
    {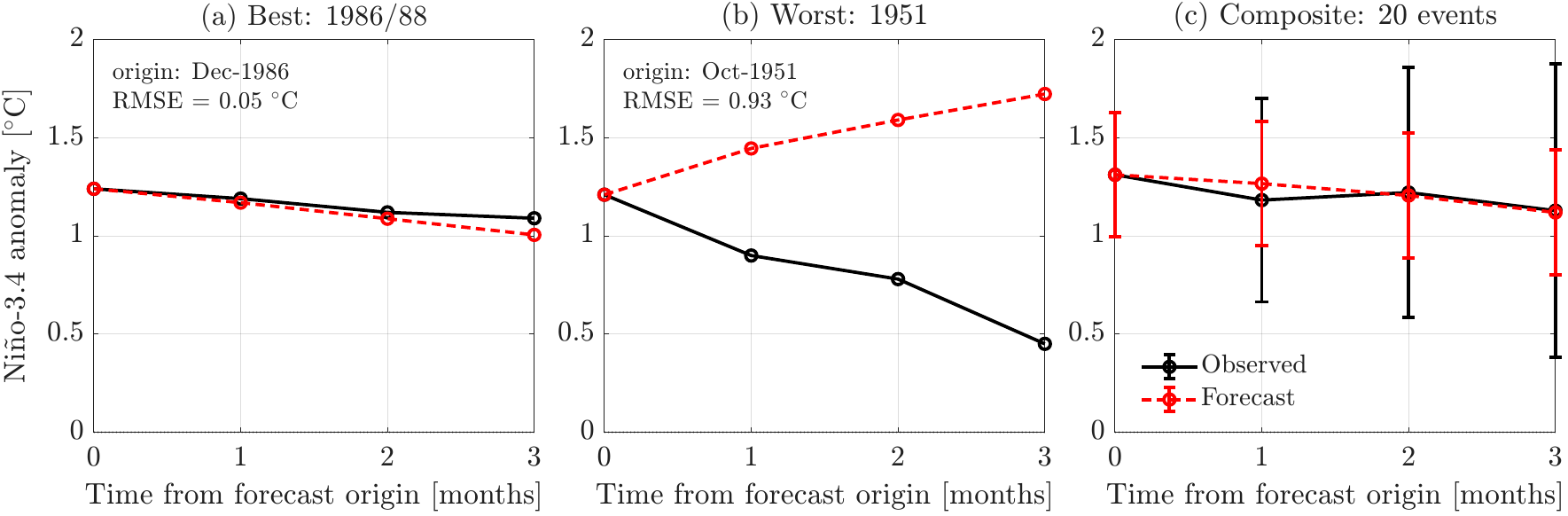}
    \caption{Historical three-month recursive SINDy hindcasts obtained by iterating the \(\Delta=1\) month map. (a) Event with the smallest hindcast RMSE. (b) Event with the largest hindcast RMSE. (c) Mean observed and forecast evolution over the 20 historical El Ni\~no events as a function of time from the forecast origin; error bars denote one standard deviation across events. Forecast origins are selected using the same delayed-state similarity criterion employed for the July 2026 initialization.}
    \label{fig:hindcasts}
\end{figure}

Figure~\ref{fig:hindcasts} illustrates the historical recursive performance of this map. The best three-month hindcast occurs for the 1986--88 event, with an RMSE of \(0.054\,^\circ{\rm C}\), whereas the largest error occurs for the 1951 event, with an RMSE of \(0.927\,^\circ{\rm C}\). Across all 20 events, the mean forecast closely follows the mean observed evolution. At recursive forecast horizons of one, two, and three months, the mean signed errors (biases) across the 20 events are \(0.084\), \(-0.015\), and \(-0.009\,^\circ{\rm C}\), respectively, while the corresponding RMSEs, which measure the magnitude of the forecast errors, are \(0.318\), \(0.489\), and \(0.650\,^\circ{\rm C}\). The recursive map therefore exhibits little systematic over- or underprediction across the historical events, although the magnitude of individual-event errors increases with forecast horizon.

The recurrent-network experiments test the same recursive problem without prescribing the temporal representation through selected delayed coordinates. GRU and LSTM networks use five-month input sequences \(N(t-4),\ldots,N(t)\), spanning the same maximum four-month history as \eqref{eq:selected-state} but additionally including \(N(t-2)\). In the pseudo-real-time historical experiments, the sequence length remains fixed while the number of available training sequences increases with the forecast date as the training record expands. Each historical hindcast remains out of sample relative to the observations used to train the corresponding network.

Among shallow GRUs with \(h=2,4,8,16,32,\) and \(64\) hidden units, historical recursive skill is best for \(h=32\), with a similarity-weighted three-month RMSE of \(0.530\,^\circ{\rm C}\) (Table~\ref{tab:recursive}). The corresponding LSTM experiment selects \(h=16\), with RMSE \(0.524\,^\circ{\rm C}\). These errors are close to the \(0.514\,^\circ{\rm C}\) obtained with the degree-one SINDy map. Allowing the temporal representation itself to be learned internally therefore provides no appreciable improvement in historical recursive skill. The best deep GRU and LSTM architectures have somewhat larger similarity-weighted errors, \(0.622\,^\circ{\rm C}\) and \(0.617\,^\circ{\rm C}\), respectively.

\begin{table}[t]
\centering
\begin{tabular}{lcc}
\toprule
Model & Selected complexity & RMSE [\(^{\circ}{\rm C}\)]\\
\midrule
SINDy     & \(p=1\)                  & 0.514\\
GRU       & \(h=32\)                 & 0.530\\
LSTM      & \(h=16\)                 & 0.524\\
Deep GRU  & \([64,32,16,8,4,2]\)    & 0.622\\
Deep LSTM & \([32,16,8,4,2]\)       & 0.617\\
\bottomrule
\end{tabular}
\caption{Historical three-month recursive model comparison. RMSE denotes the similarity-weighted error over completed historical El Ni\~no events, with greater weight assigned to forecast-origin delayed states closer to the July 2026 initialization. Recurrent-network forecasts are summarized over independent random initializations.}
\label{tab:recursive}
\end{table}

The distinction between shallow and deep recurrent networks is not uniform across events: in both families, 11 of the 20 historical events favor the selected shallow model and 9 favor the deep model. The shallow models nevertheless have lower similarity-weighted aggregate errors and perform better for five of the six historical forecast origins having the smallest delayed-state distance from the July 2026 initialization. These results favor the shallow architectures for the prospective forecast, while the deep architectures provide a useful contrast for assessing dependence on recurrent architecture.

The historical experiments reinforce the direct-forecast results. A sparse linear recurrence on explicitly selected delayed coordinates performs comparably to recurrent networks that construct their temporal representations internally, while neither polynomial nonlinearity nor increased recurrent complexity provides a systematic gain. The principal improvement in forecast skill therefore arises from representing the recent history of the index rather than from increasing the complexity of the map acting on that information.

\begin{figure}[t]
    \centering
    \includegraphics[width=\columnwidth]{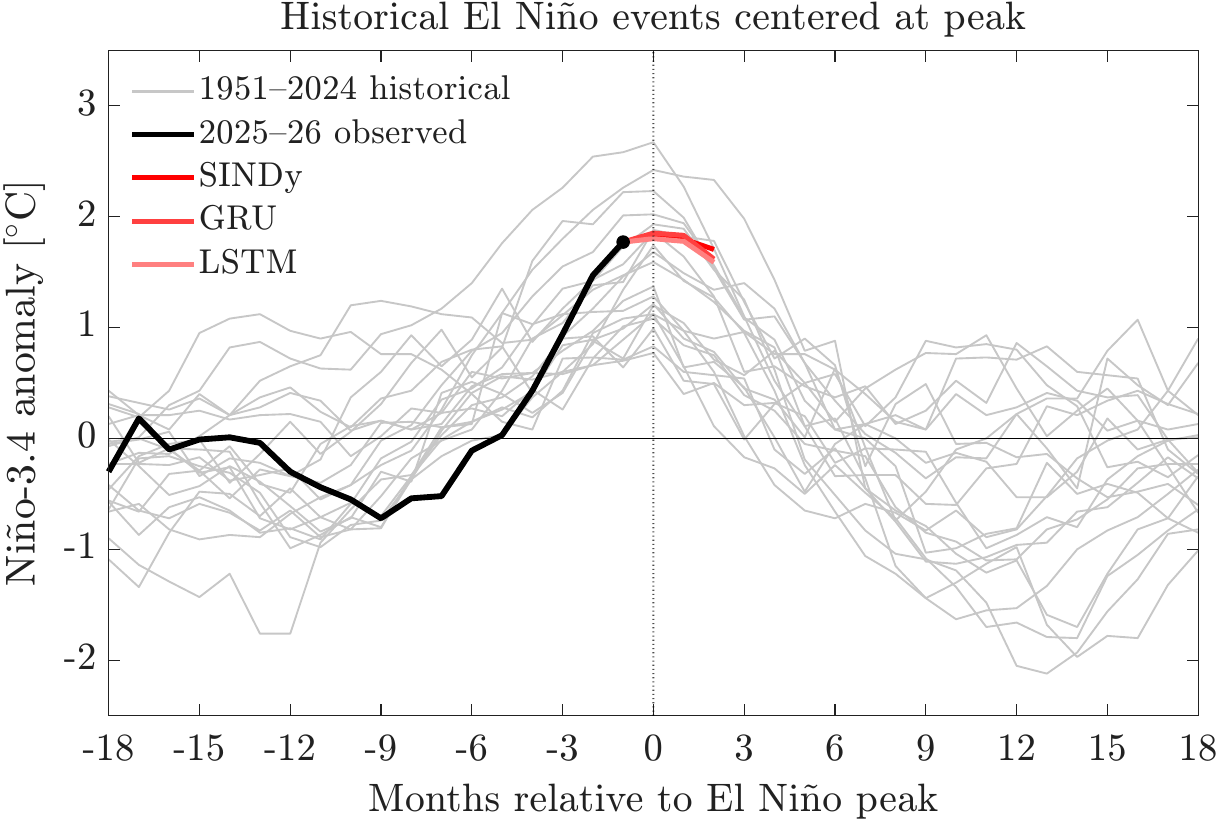}
    \caption{Prospective three-month forecasts initialized in July 2026, shown together with completed historical El Ni\~no events centered at their observed maxima (gray). The observed 2025--26 evolution is shown in black.}
    \label{fig:current}
\end{figure}

As a prospective application, the selected models are retrained using all observations available through July 2026 and iterated recursively for three months. SINDy and the selected shallow GRU and LSTM give closely agreeing forecasts, with the anomaly remaining near its July value through August--September before weakening in October. The deep recurrent models instead predict more rapid weakening, illustrating the dependence of the prospective forecast on recurrent architecture. Figure~\ref{fig:current} places the preferred forecasts in the context of the 20 completed historical El Ni\~no events; among those events, the predicted evolution is most similar to the 1965--66 El Ni\~no.

These results indicate that much of the forecast-relevant information in the Ni\~no-3.4 record is captured by a small number of past monthly observations. Direct forecast skill saturates with a compact delayed representation, and an MLP provides essentially no improvement over ridge regression on the same coordinates. SINDy likewise provides no systematic evidence that polynomial nonlinearity improves direct forecast skill, while historical recursive tests favor a linear recurrence. GRU and LSTM networks, which instead learn their temporal representations internally, provide no appreciable improvement in historical recursive skill, while deep architectures have larger aggregate historical errors than their shallow counterparts. The consistency across these distinct model classes supports a compact predictive representation of Ni\~no-3.4 evolution over the short-to-intermediate forecast leads and recursive horizons considered here.

\section{Conclusions}
\label{sec:conclusions}

The Ni\~no-3.4 record contains substantial predictive information in a small number of delayed observations. Ridge regression shows that this information saturates with a compact set of coordinates spanning short and longer memory scales, while MLP and SINDy models show no systematic gain from additional nonlinear complexity. Historical recursive experiments show that GRU and LSTM networks that learn their temporal representations internally do not improve appreciably on the explicit delayed-coordinate SINDy model.

Recursive robustness tests favor a simple linear SINDy recurrence over higher-degree polynomial maps. Pseudo-real-time hindcasts select shallow GRU and LSTM architectures, which have lower similarity-weighted historical errors than the selected deep architectures. As a prospective application, the SINDy and shallow recurrent models are applied to the developing 2026 event and give closely agreeing short-term forecasts, while the deep recurrent models illustrate the sensitivity of this forecast to recurrent architecture. Over the direct forecast leads and recursive forecast horizons considered here, a compact representation of the recent history of Ni\~no-3.4 appears more consequential for prediction than increasing model complexity.

\section*{Acknowledgments}

This work was initiated during the workshop \emph{Oceanograf\'ia en un Clima Cambiante: Procesos F\'isicos, Biol\'ogicos y Pesqueros desde la Teor\'ia, la Observaci\'on y la Modelaci\'on}, held at Universidad del B\'io-B\'io, Concepci\'on, Chile, in August 2026. The author gratefully acknowledges the hospitality of Universidad del B\'io-B\'io and the workshop organizers.

\section*{Conflict of Interest}

The author declares no conflicts of interest.

\section*{Availability Statement}

The monthly Ni\~no-3.4 anomaly record used in this study is publicly available from the NOAA Climate Prediction Center at \url{https://www.cpc.ncep.noaa.gov/data/indices/detrend.nino34.ascii.txt}. MATLAB codes used for the analysis are available from the author upon reasonable request.

\bibliographystyle{apalike}
\bibliography{fot}

\end{document}